# Inertial migration of bidisperse suspensions flowing in microchannels: effect of particle diameters ratio

T. Tohme*[1], Y. Gao[1], P. Magaud[1, 2], L. Baldas[1], C. Lafforgue[3], S. Colin[1]

[1]Institut Clément Ader (ICA), Université de Toulouse, INSA, ISAE-SUPAERO, Mines-Albi, UPS, 3 rue Caroline Aigle, 31400 Toulouse, France
[2]Université de Limoges, 33 rue François Mitterrand, 87032 Limoges, France
[3]Laboratoire d'Ingénierie des Systèmes Biologiques et des Procédés (LISBP), Université de Toulouse, CNRS, INRA, INSA, 135 Avenue de Rangueil, 31077 Toulouse, France
(E-mail: tohme.tohme@univ-tlse3.fr; pascale.magaud@insa-toulouse.fr; lucien.baldas@insa-toulouse.fr; christine.lafforgue@insa-toulouse.fr; stephane.colin@insa-toulouse.fr)

**Abstract**
Up to date, inertial migration of particles in microflows has demonstrated a great potential for a wide range of applications. In particular, this phenomenon is used to achieve particle separation or sorting in a suspension. Recent works reported that the focusing mode of particles can be modified in a polydisperse suspension. Nevertheless, the impact of the particle sizes in a mixture on their inertial migration has been rarely studied up to now. Thus, we have investigated in this work the influence of bidispersity on the lateral migration of the particles towards equilibrium positions and on their longitudinal ordering into trains. Different changes in the particles behavior were observed when the ratio between the particle sizes ($d_{p1}/d_{p2}$) varied from 1.64 to 4.58.

## INTRODUCTION
Poiseuille (1836) was the first one to note the inhomogeneous radial distribution of blood corpuscles in a capillary vessel. Since then, it is well known that the distribution of particles flowing in a channel can become inhomogeneous even in very simple flow and geometrical configurations. Following the observations of Poiseuille, many investigators tackled the problem of inertial migration numerically and experimentally. Di Carlo *et al*. (2007) showed that, in square channels, particles focus at moderate Reynolds numbers towards four equilibrium regions centered at the faces of the channels. Matas *et al*. (2004) observed that particles flowing in micro-channels not only laterally migrate towards specific positions but although longitudinally order to form trains. The great potential of inertial migration is now obvious and several applications have already been developed such as the separation of particles from their environment. However, the majority of the to-date studies were done on monodisperse suspensions (Abbas *et al*. 2014), whereas most of the possible applications concern polydisperse suspensions. The aim of the present work is thus to better understand the inertial migration phenomenon in a model bidisperse suspension. Both lateral migration and longitudinal ordering are experimentally studied in square micro-channels for different flow conditions.

## EXPERIMENTAL METHOD
Model neutrally buoyant particles were polystyrene microspheres ($\rho$ = 1050 kg/m$^3$) of different diameters $d_p$: 1.9 µm ($d_p/H$ = 0.02), 5.3 µm ($d_p/H$ = 0.066), 8.7 µm ($d_p/H$ = 0.11) and 15.6 µm ($d_p/H$ = 0.195). Square borosilicate channels with an inner height $H$ = 80 µm and a length of 10 cm were used. Three bidisperse suspensions were prepared with different ratios of size ($d_{p1}/d_{p2}$ = 1.64, 2.94 and 4.58). An experimental set-up based on classical microscopy was used to visualize *in situ* the suspensions flowing in the micro-channels. Sequences of 2000 top-view images were taken in the midplane of the micro-channel. Positions of the different-size particles were determined from image post-processing. Since in square micro-channels particles are supposed to migrate to four equilibrium positions located near the channel face centres, a so-called "focusing degree" $\eta$, which corresponds to "the percentage of particles focused at the four equilibrium positions in comparison with the total number of particles in the microchannel" is defined (see (Gao *et al*. 2017) for more details).

## RESULTS
Figure 1.a. and b. compare the focusing degree for bidisperse suspensions with respectively a low $d_{p1}/d_{p2}$ ratio (8.7-µm and 5.3-µm particles) and a high $d_{p1}/d_{p2}$ ratio (15.6 µm and 5.3 µm) and the focusing degree observed for each type of particles in monodisperse suspensions.

---
* Corresponding author

In the case of low particle size ratios (Figure 1.a.), the focusing degree of both types of particles in monodisperse and bidisperse suspensions first increases with *Re* up to a maximum around *Re* = 110, suggesting that the distance between the measurement zone and the channel inlet is shorter than the streamwise length required for focalization. The decrease of the focusing degree observed for higher Reynolds numbers could be due to the transition towards new additional equilibrium positions located near the channel corners which was numerically and experimentally shown by Nakagawa *et al.* (2015) for the same Reynolds range. Focusing degree profiles for the bidisperse suspension are quite similar to those obtained for monodisperse suspensions. No influence of the bidispersity on the lateral migration of the two types of particles is thus evidenced. Both sizes of particles migrate towards the same equilibrium positions.

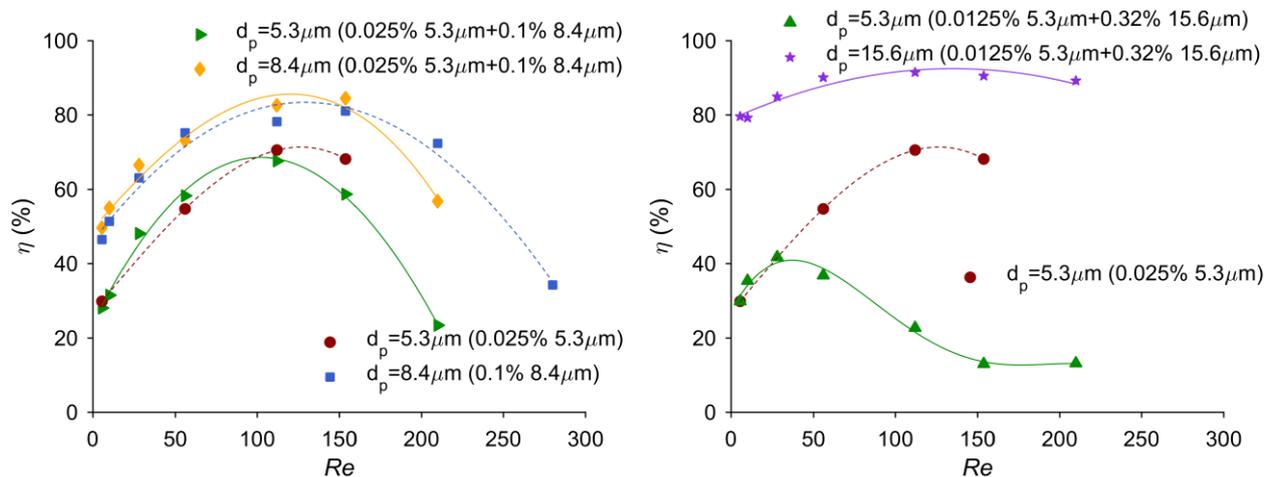

**Figure 1.** Focusing degree versus *Re* in monodisperse (dotted lines) and bidisperse suspensions (solid lines) for **a.** a low $d_{p1}/d_{p2}$ ratio: 8.7-µm and 5.3-µm particles **b.** a high $d_{p1}/d_{p2}$ ratio: 15.6 µm and 5.3 µm. Volume concentrations of each particle type are indicated into brackets.

However, in the case of a high $d_{p1}/d_{p2}$ ratio, the focusing degree for the small particles in the bidisperse suspension is much lower than that observed in the monodisperse one (Figure 1.b.). The high level (> 80%) of the focusing degree of the large particles in the bidisperse suspension indicates that the four equilibrium positions are mainly occupied by these particles. As a consequence, the small particles remain concentrated on an annulus close to the channel walls, as confirmed by our observations, which leads to a low value of the focusing degree for these particles. The bidispersity thus affects drastically the lateral migration of the small particles and the two types of particles do not concentrate in the same regions, which could facilitate their separation or their sorting.

## CONCLUSION
In this work, the effect of bidispersity on the inertial focusing behaviors of mixed suspensions of particles is presented. When the ratio between the particle sizes ($d_{p1}/d_{p2}$ = 1.64) is close to one, bidispersity is found to have little effect on the particle migration. However, for high ratios (i.e., $d_{p1}/d_{p2}$ = 4.58), the bidispersity highly affects the lateral migration of the small particles and the separation thus seems to be possible. Further experiments need to be conducted to confirm these results and extend them to other sizes of particles and other experimental conditions.

## AKNOWLEDGMENTS
The authors acknowledge the FERMaT research federation for its support.